\documentclass[10pt,twocolumn,english,twocolumn]{revtex4-1}
\usepackage{amssymb}
\usepackage{graphicx}

\makeatletter

\usepackage{lmodern}
\usepackage[T1]{fontenc}

\makeatother

\usepackage{babel}
\begin{document}

\title{Counting the Cycles of Light using a Self-Referenced Optical Microresonator}

\author{J. D. Jost\textsuperscript{1,{$\dagger$}}, T. Herr\textsuperscript{1,2,{$\dagger$}},
C. Lecaplain\textsuperscript{1}, V. Brasch\textsuperscript{1}, M.
H. P. Pfeiffer\textsuperscript{1}, T. J. Kippenberg\textsuperscript{1}}
\email{tobias.kippenberg@epfl.ch}

\affiliation{1. \'{E}cole Polytechnique F\'{e}d\'{e}rale de Lausanne (EPFL), CH-1015 Lausanne,
Switzerland.\\2. Centre Suisse d'Electronique et de Microtechnique
(CSEM), Ne\^{u}chatel, Switzerland }


\begin{abstract}

Phase coherently linking optical to radio frequencies
with femtosecond mode-locked laser frequency combs enabled counting
the cycles of light and is the basis of optical clocks, absolute frequency
synthesis, tests of fundamental physics, and improved spectroscopy.
Using an optical microresonator frequency comb to establish a coherent
link between optical and microwave frequencies will extend optical frequency synthesis and measurements to areas requiring
compact form factor, on chip integration and comb line spacing in
the microwave regime, including coherent telecommunications, astrophysical
spectrometer calibration or microwave photonics. Here we demonstrate a microwave to optical link with a microresonator.
Using a temporal dissipative single soliton state in an ultra-high
Q crystalline microresonator that is broadened in highly nonlinear fiber an optical frequency comb is generated
that is self-referenced, allowing to phase coherently link a 190 $\mathrm{THz}$
optical carrier directly to a 14 $\mathrm{GHz}$ microwave frequency.
Our work demonstrates precision optical frequency measurements
can be realized with compact high Q microresonators.
\end{abstract}

\maketitle

The development of the optical frequency comb (OFC) based on femtosecond pulsed mode-locked lasers
\cite{Jones2000,Diddams2000,Ye2003} in conjunction with nonlinear spectral
broadening constituted a dramatic simplification over large scale
harmonic frequency chains \cite{Evenson1973}. In the frequency domain
OFCs give an equidistant optical lines, where the frequency of each component
obeys $f_{n}=n\cdotp f_{rep}+f_{0}$. The spacing between the
lines is determined by the pulse repetition rate of the laser $f_{rep}$
($n$ being an integer). Knowledge of the comb's overall offset frequency
$f_{0}$ (also referred to as the carrier envelope offset frequency) along
with $f_{rep}$ and $n$ allows linking the optical frequency $f_{n}$
to the electronically countable frequency $f_{rep}$ and $f_{0}$
in the radio frequency or microwave domain. In this way the measurement
of $f_{rep}$ and $f_{0}$ corresponds to counting the cycles of light.
Self-referencing of the comb, i.e. a self-contained measurement of
$f_{0}$ and $f_{rep}$, has been achieved using nonlinear interferometers,
whereby the combs bandwidth needs to be broadened to encompass typically
two-thirds of or a full octave \cite{Reichert1999,Telle1999,Morgner2001,Diddams2003a}.
This measurement is a key prerequisite for many applications of OFCs.
For example, referencing one of the comb lines of an OFC to an atomic
frequency standard \cite{Diddams2001a} allows the comb to function
as a 'gearbox' realizing the next generation of atomic clocks based
on optical transitions. Self-referenced frequency combs can also be
used for optical frequency synthesis, and have enabled the most accurate
frequency measurements \cite{Rosenband2008,Hinkley2013}. Frequency combs with large mode spacings ($\geq$ 10 GHz), which are challenging to obtain via mode-locked lasers \cite{Bartels2009a}, can be used in a growing number of applications, including astronomical spectrometer calibration \cite{Steinmetz2008}, dual comb coherent Raman imaging \cite{Ideguchi2013}, high speed
optical sampling or coherent telecommunications \cite{Pfeifle2014}.
In each of these applications, having a line spacing $\gtrsim10$ $\mathrm{GHz}$ is either beneficial or even required.


\begin{figure*}[t]
\centering\hspace*{-1ex}\includegraphics[width=.9\linewidth]{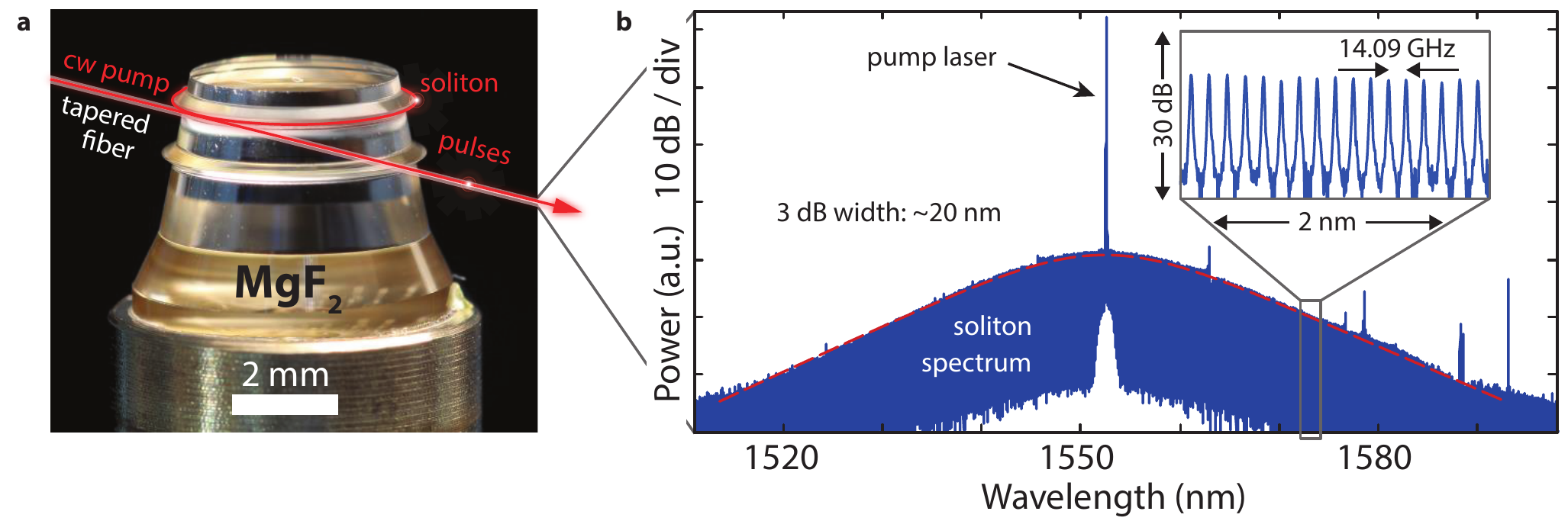}\caption{\label{fig:Microresonator-and-Soliton}
Crystalline $\mathrm{MgF_{2}}$
microresonator and temporal dissipative soliton generation:\emph{
a.} Optical image of the employed ultra high Q crystalline whispering
gallery optical microresonators on a magnesium fluoride pillar with
a diameter of several millimeters. The ultra high Q whispering gallery
optical modes are confined in the fabricated protrusions that extend
around the circumference. The top resonator was used in the experiments
and the mode of interest has a free spectral range of $14.0939\:\mathrm{GHz}$
and a loaded $Q\approx10^{9}$. \emph{b.} The hyperbolic-secant shaped
spectrum (fit: red dotted line) of the single temporal soliton produced
inside the resonator by the continuous wave pump laser. The inset
shows the ability to resolve the microresonator comb lines on a grating
based spectrometer. }
\end{figure*}

Different from mode-locked lasers, microresonator
frequency combs (MFC) \cite{DelHaye2007,Kippenberg2011} are generated using parametric frequency conversion
\cite{Kippenberg2004a,Savchenkov2004}, of a continuous wave (CW)
laser. This approach exhibits several attractive features that have
the potential to extend further the use of OFCs to new areas in precision
optical measurement, spectroscopy, astronomy, telecommunications and
industrial applications. Fundamentally different from mode-locked laser
systems, MFCs offer: high repetition rates ($>10$ $\mathrm{GHz}$),
compact form factor, broadband parametric gain that can be generated
in wavelength regimes ranging from the visible \cite{Savchenkov2011} to the mid-infrared \cite{Wang2013, Griffith2015}, CMOS compatible microresonator platforms,
and high power per comb line. A unique characteristic of MFCs is the pump laser constitutes one of the frequency comb
components and there is no active gain laser medium in the system. These
properties have encouraged in recent years the intense investigation
of microresonator based frequency combs with the ultimate goal of
realizing a self-referenced system. Progress in recent years includes new microresonator
platforms in crystalline materials \cite{Savchenkov2008c}, fused
silica microtoriods \cite{DelHaye2007}, and photonic chips (based
on Silicon Nitride \cite{Foster2011a,Levy2010}, Aluminum Nitride \cite{Jung2013}, Hydex$^{\circledR}$
\cite{Peccianti2012,Razzari2010} and diamond \cite{Hausmann2014}).
In addition, MFCs without self-referencing
have been used for coherent telecommunications \cite{Pfeifle2014},
compact atomic clocks \cite{Papp2014}, stabilized oscillators \cite{Savchenkov2013a},
and optical pulse generation \cite{Ferdous2011,Saha2012a,Herr2013}.
The dynamics of microresonator frequency combs have been investigated,
and regimes with low noise frequency comb operation have been identified
based on intrinsic low phase noise regimes, or via tuning mechanisms
such as $\delta-\Delta$ matching \cite{Herr2012}, parametric seeding
\cite{DelHaye2014}, injection locking \cite{Li2012,DelHaye2014}
or via the observation of phase-locking \cite{Saha2012a}. Moreover
recently, low noise frequency combs have been generated via temporal dissipative soliton formation \cite{Wabnitz1993,Leo2010,Grelu2012,Herr2013} and numerical tools to simulate comb dynamics emerged based on the Lugiato-Lefever equation \cite{Lugaito1987,Coen2013,Lamont2013} and the coupled-modes equation\cite{Chembo2010,Herr2013}.

However, despite the rapid progress
in understanding, simulations, applications and platforms, an outstanding
milestone has not been reached: a self-referenced microresonator system,
capable of phase coherently linking the microwave and the optical frequency
domain. So far, knowledge of the absolute frequency of all comb lines of a MFC has only been achieved using an auxiliary self-referenced fiber laser frequency comb as a reference \cite{DelHaye2008}, as well as via locking of two comb teeth to a Rubidium transition \cite{Papp2014}. Self-referencing
however, has never been achieved. One reason for this is the high
repetition rate of the microresonators leading to correspondingly
low pulse peak intensities. This makes external spectral broadening using nonlinear
fiber, which is widely employed in mode-locked
lasers, difficult to apply. While octave spanning combs \cite{DelHaye2011,Okawachi2011}
have been attained directly from MFCs, they have not been suitable
for self-referencing techniques due to excess noise associated with
subcomb formation \cite{Herr2012}.

Here, we demonstrate a coherent microwave to optical
link using temporal dissipative soliton formation in a crystalline
microresonator in conjunction with external spectral broadening, measuring
simultaneously both $f_{rep}$ and $f_{0}$ that are necessary for
linking the optical to the radio frequency domain. Our approach uses the newly discovered class
of temporal dissipative cavity solitons in microresonators \cite{Akhmediev1987,Akhmediev2008,Leo2010,Herr2013}
and marks the first time that the carrier envelope offset frequency
of such a soliton has been measured. Our results demonstrate that MFCs generated by soliton formation are suitable for absolute frequency measurements. In particular the pulse to pulse timing jitter is sufficiently low to enable precise measurement of the carrier envelope offset frequency via self-referencing. Although not demonstrated here, it has already been
shown that microresonator comb parameters $f_{rep}$ and $f_{0}$
can be stabilized by controlling the pump laser frequency and by actuating the resonator free spectral
range by heating or mechanical stress \cite{DelHaye2008,Papp2012}. 


\begin{figure*}
\centering{}
\includegraphics[width=1\linewidth]{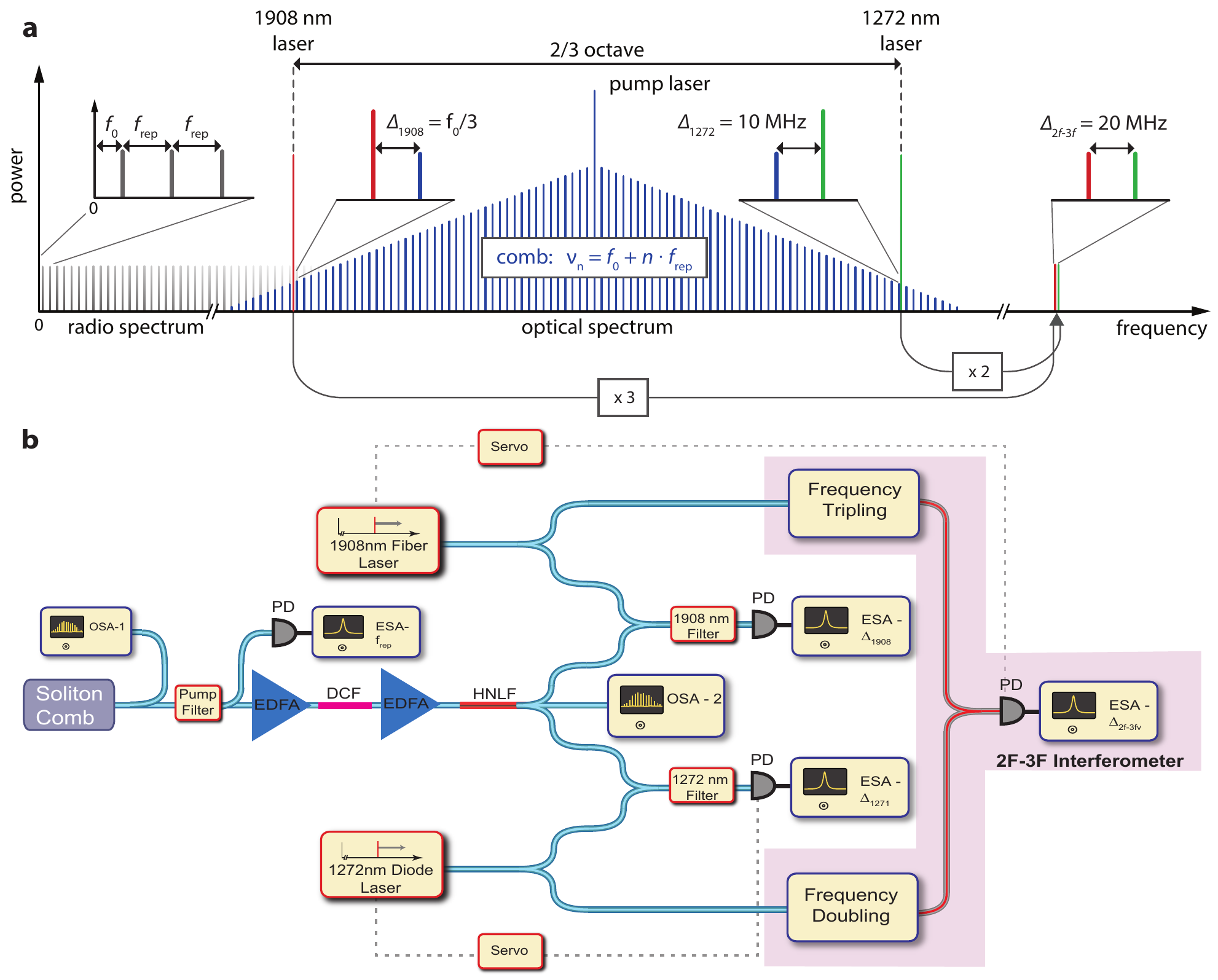}
\caption{\label{fig:Experimental-Setup-and}Experimental setup
and $2f-3f$ microresonator self-referencing scheme:\emph{ a. }The
frequency domain picture showing the relevant frequency components
used to self-reference the comb and to determine the carrier envelope
offset frequency ($f_{0}$). \emph{b.} The simplified experimental
setup used for self-referencing. A portion of the solitons that are
outcoupled from the resonator are sent to an OSA to measure the spectrum, then residual pump light is filtered out using fiber optic filters before a portion is picked off 
and sent to a photodetector (PD) to measure the repetition rate on an electronic
spectrum analyzer (ESA). The pulse is first preamplified in an erbium doped fiber amplifier (EDFA) and then prechirped to temporally
broaden it with a dispersion compensating fiber (DCF) before being
amplified by a high power EDFA. The pulse is subsequently recompressed
and coupled into a highly nonlinear fiber (HNLF) where the coherent
supercontinuum is generated. A fraction of the spectrum is mixed with
light from the $1908$ $\mathrm{nm}$ Thulium fiber laser and sent
through a $1908$ $\mathrm{nm}$ bandpass filter to a PD and ESA to
measure $\Delta_{1908}$. The same is done with the $1272$ $\mathrm{nm}$
external cavity diode laser to measure $\Delta_{1272}$ and a servo
loop is used to phase lock the laser to the optical frequency comb
and fix $\Delta_{1272}$ where a signal from an atomic clock is used
as a reference. To create the $2f-3f$ interferometer light from the
$1272$ $\mathrm{nm}$ laser is frequency doubled in a periodically
poled Lithium Niobate crystal (PPLN) to produce light at $636$ $\mathrm{nm}$.
Light from the $1908$ $\mathrm{nm}$ laser is frequency doubled to
$954$ $\mathrm{nm}$ in a PPLN crystal, and subsequently combined
with $1908$ $\mathrm{nm}$ and sent through a PPLN crystal phase
matched for sum frequency generation creating light at $636$ $\mathrm{nm}$.
The generated visible light is optically heterodyned on a PD with
the frequency doubled light from the $1272$ $\mathrm{nm}$ laser,
permitting to measure $\Delta_{2f-3f}$ on an ESA. This offset frequency
is fixed by phase locking the $1908$ $\mathrm{nm}$ laser via the
$2f-3f$ interferometer. With this scheme the carrier envelope frequency
is measured by recording $\Delta_{1908}$. }
\end{figure*}


Optical microresonators support different azimuthal optical whispering
gallery modes (WGM). The free spectral range (FSR) between modes of a particular mode family is determined
by material and geometric dispersion. It has been shown that WGM
can have quality factors exceeding $10^{10}$ \cite{Braginsky1989,Grudinin2006}.
When a CW pump laser is coupled to a WGM the resonator's Kerr nonlinearity can lead to the efficient nonlinear frequency conversion. The
resonator used in this work is a crystalline ultra high Q resonator
made by polishing crystalline $\mathrm{MgF_{2}}$ \cite{Hofer2010,Ilchenko2004,Herr2014}
(cf. figure \ref{fig:Microresonator-and-Soliton}), and can support
WGMs \cite{Braginsky1989} confined in one of
its protrusions that extend around the circumference of
the resonator. The mode used has a quality factor of $\sim10^{9}$
and a free spectral range of $14.0939$ GHz. Light from a continuous
wave fiber laser at $1553$ $\mathrm{nm}$ with $\sim150$ mW can
be coupled into and out of the resonator via evanescent coupling using
a tapered optical fiber \cite{Spillane2003}. To form the solitons
in the cavity, the pump laser's frequency is scanned over the resonance
and stopped when the appropriate conditions are met \cite{Herr2013}.
The duration of the pulse inside the resonator depends mainly on the coupling
and detuning of the pump laser to the resonator mode, and can be estimated
from the bandwidth of the spectrum shown in figure \ref{fig:Microresonator-and-Soliton},
to be $\sim130$ fs. The optical spectrum generated by the soliton
is not yet sufficiently broad for self-referencing; however, the spectrum
can be broadened via supercontinuum generation \cite{Dudley2006}.

\begin{figure}[h]
\centering\hspace*{-3ex}
\includegraphics[width=.9\linewidth]{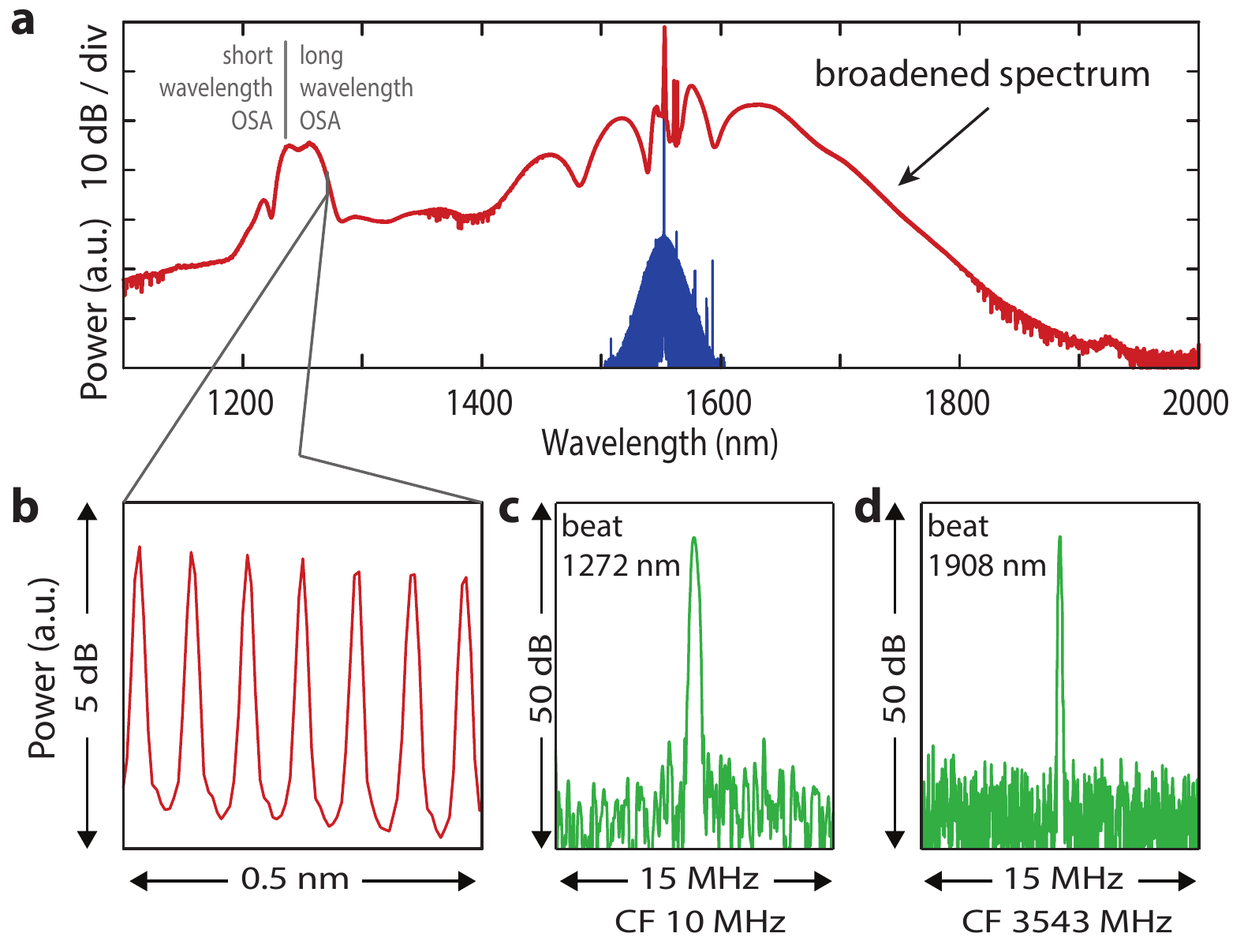}
\caption{\textbf{\label{fig:Optical-Spectrum:-The}}Optical spectrum of the
microresonator before and after external broadening: \emph{a.} The
blue trace shows the optical spectrum generated in the crystalline
optical microresonator by the temporal dissipative soliton state.
The large central spike originates from residual light from the pump
laser. The spectrum after the supercontinuum generation is denoted
in red. It is composed of data take from two different OSAs as result
of the limited bandwidth of the individual instruments. \emph{b.}
Zoom into the broadened spectrum revealing the widely spaced comb
lines. \emph{c.} Heterodyne beatnote of a laser at 1272 $\mathrm{nm}$
with the broadened comb demonstrating a signal to noise ratio exceeding
40 dB in the resolution bandwidth (RBW) of 300 $\mathrm{kHz}$. \emph{d.}
shows heterodyne beatnote at the long wavelength end of the comb at
1900 $\mathrm{nm}$ (RBW 100 $\mathrm{kHz}$).}
\end{figure}

The experimental setup after the resonator is shown in figure \ref{fig:Experimental-Setup-and}. A portion of the pulse train produced by the resonator is sent to an optical spectrum analyzer (OSA). The rest of the pulse train then has the CW background consisting of
the residual pump laser \cite{Herr2013} minimized by fiber optic filters. Then another small amount is sent to a photodetector and a electronic spectrum analyzers (ESA) to detect $f_{rep}$. The pulse is then preamplified with an erbium doped fiber amplifier (EDFA). The main type of fiber in the experiment is SMF28, which has anomalous group velocity
dispersion in the wave length range of the soliton. Before being sent
into a high EDFA the soliton
pulse is prechirped using dispersion compensating fiber (DCF), which
has normal group velocity dispersion, to minimize nonlinear effects
in the EDFA. In this way the average power of the pulse train is increased to 2 W. After the EDFA, the pulse
is recompressed using SMF28 fiber via the cut back method to a duration
of $\sim300$ fs with an energy of $~\sim150$ $\mathrm{pJ}$. This
pulse is subsequently sent through approximately $2\,\mathrm{m}$
of highly nonlinear fiber (HNLF)(Menlo Systems), where supercontinuum generation
occurs. The resulting spectrum can be seen in figure \ref{fig:Optical-Spectrum:-The}.
The blue trace shows the soliton spectrum after the optical microresonator,
and the red after the HNLF fiber. The latter spectrum exceeds two-thirds
of an octave, which is sufficient for self-referencing via a $2f-3f$
interferometer \cite{Reichert1999,Telle1999,Morgner2001,Diddams2003a}.
Importantly, the broadened spectrum is coherent. This is verified
by using a heterodyne beat with additional external lasers at the two ends of the comb, as detailed
below. Figure \ref{fig:Optical-Spectrum:-The}b shows a zoom into the
spectrum taken after the HNLF fiber where the individual comb lines
are clearly visible, even with the limited resolution of the OSA ($0.02$
$\mathrm{nm}$). 


\begin{figure}[!htbp]
\centering\hspace*{-2ex}
\includegraphics[width=0.47\textwidth]{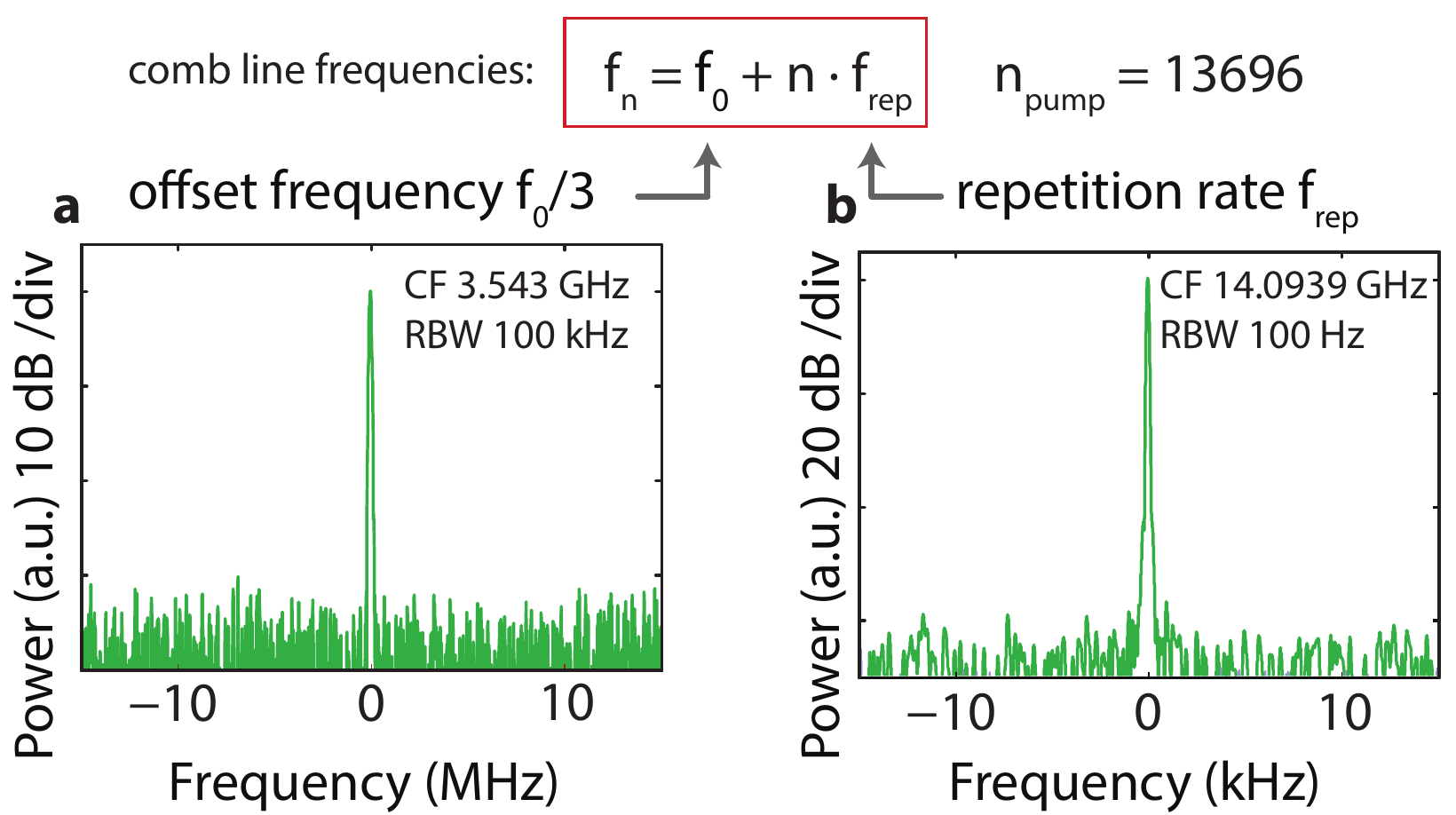}
\caption{\label{fig:Self-referencing-signals:A.-The}Repetition
rate and carrier envelope frequency signals of the self-referenced
microresonator comb:\emph{ a.} The offset frequency ($f_{0}$) of
the optical microresonator frequency comb divided by three as described
in eqn. \ref{eq:delta1908=00003Dfo}. The measured optical heterodyne
beat frequency has a center frequency of $3.543$ $\mathrm{GHz}$
and exhibits a signal to noise that exceeds 30 dB in a 100 $\mathrm{kHz}$
RBW. \emph{b.} The repetition rate $f_{rep}$ of the soliton in the
optical microresonator with a center frequency (CF) of $14.0939$
$\mathrm{GHz}$ and a signal to noise ratio (SNR) $>$ 60 dB measured
in a resolution bandwidth of 100 $\mathrm{Hz}$. The large SNRs are sufficient for accurate phase tracking of the two microwave signals.}
\end{figure}


Self-referencing is achieved by measuring $f_{rep}$ and $f_{0}$
of the generated frequency comb. By picking off a small portion of
the light after it leaves the resonator and sending it to a photodetector
$f_{rep}$ can be directly measured (cf. figure \ref{fig:Experimental-Setup-and}
). The broadened spectrum is sufficiently wide and allows employing
a $2f-3f$ interferometer \cite{Reichert1999} to determine $f_{0}$.
Traditionally this is implemented by frequency tripling a component
of the low frequency part of the spectrum $f_{L}=n\cdotp f_{rep}+f_{0}$
using a combination of second harmonic and sum frequency generation to give $3f_{L}=3n\cdotp f_{rep}+3f_{0}$,
where $n$ is an integer. In addition, a component of the higher frequency
part of the spectrum $f_{H}=m\cdotp f_{rep}+f_{0}$ is frequency doubled
using second harmonic generation to give $2f_{H}=2m\cdotp f_{rep}+2f_{0}$
where $m$ is an integer. Mixing the doubled and the tripled light
and detection on a photodetector gives the offset frequency $3f_{L}-2f_{H}=f_{0}$
, granted $3n=2m$, i.e. necessitating a spectrum that covers two-thirds
of an octave. Here a scheme involving two transfer lasers is implemented,
which has the advantage of allowing independent verification of the
coherence of the supercontinuum generation at the two ends of the
spectrum. The coherence of the generated broadband OFC is verified by optically
heterodyning the two reference lasers at $\sim1272$ $\mathrm{nm}$
(external cavity diode laser) and $\sim1908\,\mathrm{nm}$ (Thulium
fiber laser) with the supercontinuum and detecting the optical heterodyne
beat signal on a photodetector (see figure \ref{fig:Optical-Spectrum:-The}).
The beatnote frequencies can be written in terms of the frequency comb parameters
and an offset as
\begin{equation}
f_{1272}=n\cdotp f_{rep}+f_{0}+\Delta_{1272}
\end{equation}
and
\begin{equation}
f_{1908}=m\cdotp f_{rep}+f_{0}-\Delta_{1908}
\end{equation}
where $n$ and $m$ are integers and $\Delta_{1272}$ and $\Delta_{1908}$
are the frequency offsets ($>0$) of the transfer lasers from the
generated OFC (see figure \ref{fig:Optical-Spectrum:-The}). The offset
frequency $\Delta_{1272}$ and $\Delta_{1908}$ are chosen to be both positive by convention. The $2f-3f$ interferometry is constructed with the
reference lasers where one arm of the interferometer serves for second
harmonic generation of the light at $1272$ $\mathrm{nm}$ to give
light at $636$ $\mathrm{nm}$ written as (see figure \ref{fig:Experimental-Setup-and})
\begin{equation}
2f_{1272}=2n\cdotp f_{rep}+2f_{0}+2\Delta_{1272}.
\end{equation}
The other interferometer arm serves for frequency tripling of the light
at $1908$ $\mathrm{nm}$ via second harmonic generation to create
light at $954$ $\mathrm{nm}$ followed by sum frequency generation
of the $954$ $\mathrm{nm}$ and $1908$ $\mathrm{nm}$ light to give
light at $636$ $\mathrm{nm}$, and the frequency can be written as:
\begin{equation}
3f_{1908}=3m\cdotp f_{rep}+3f_{0}-3\Delta_{1908}.
\end{equation}
The doubled and tripled light are mixed and detected on a photodetector,
giving an optical heterodyne signal at the difference frequency:
\begin{equation}
\Delta_{2f-3f}=2f_{1272}-3f_{1908}.
\end{equation}
The offset frequency $f_{0}$ is related to the beats of the transfer lasers
with the generated frequency comb: 
\begin{equation}
\Delta_{2f-3f}=\left(2n-3m\right)f_{rep}-f_{0}-3\Delta_{1908}+2\Delta_{1272}.
\end{equation}
The transfer laser at $1272$ $\mathrm{nm}$ is phase locked to a
frequency comb component with $\Delta_{1272}=10$ MHz offset frequency.
The frequency tripled $1908$ $\mathrm{nm}$ transfer laser is phase locked
via the $2f-3f$ interferometer signal to $20$ $\mathrm{MHz}$ below
the the frequency doubled $1272$ $\mathrm{nm}$ transfer laser at $\Delta_{2f-3f}=20$
$\mathrm{MHz}$. Both phase locks are referenced to a commercial atomic
clock. For $2n-3m=0$ (which can readily be achieved by locking the
$1272$ $\mathrm{nm}$ transfer laser to the appropriate comb line
) the beat $\Delta_{1908}$ between the 1908 nm transfer laser and
the OFC corresponds to: 
\begin{equation}
\Delta_{1908}=\frac{f_{0}}{3}\label{eq:delta1908=00003Dfo}
\end{equation}
With this measurement technique an offset frequency of $\frac{f_{0}}{3}=3.543$ GHz (see figure \ref{fig:Self-referencing-signals:A.-The}). The signal to noise ratio (SNR) of $f_{0}$ of >30 dB
in 100 kHz resolution bandwidth (RBW), as well as a SNR of >60 dB in 100 Hz RBW of $f_{rep}$, is sufficient for accurate, i.e. real time counting of the cycles of the two radio frequency beats (and making the use of e.g. tracking
oscillators unnecessary). This, along with the knowledge of the comb teeth number, provides the
ability to directly count the cycles of the pump laser as well as the other comb teeth.

It should be noted, that the determination of the comb line index
with a mode-locked laser frequency comb with
repetition rates $<1$ $\mathrm{GHz}$, can be challenging. Here
the much denser comb spectrum does normally not permit resolving the
comb line with a grating based OSA, and moreover, the sensitivity
to drifts in repetition rate is significantly higher, necessitating
to fully phase stabilize the comb in order to determine the comb line
index. In contrast, the comb line index can be obtained in the present
work with a low resolution wavemeter without requiring locking of
either $f_{rep}$ or $f_{0}$. In this way the comb line number of
the pump laser was determined to be $n_{pump}=13696$. Self referencing
along with knowledge of the comb line numbers, implies that the absolute
laser frequency of the pump laser ($f_{pump}$, as well as any other
comb teeth) can be directly determined via the repetition rate beatnote
$(f_{rep})$ and the recorded beatnote $(\Delta_{1908})$, since the
absolute laser frequency is related to the two quantities via 
\begin{equation}
f_{pump}=3\Delta_{1908}+n_{pump}\cdot f_{rep},
\end{equation}
establishing the ability to count the cycles of light via the crystalline
microresonator.

Our results constitute the first phase coherent link from the microwave
to optical domain using a microresonator, by demonstrating measurement
of the carrier envelope frequency of a temporal dissipative soliton in a microresonator. These results demonstrate microresonator based frequency combs can provide accurate and precise absolute optical frequency standards for a wide range of applications in optical frequency metrology, optical
atomic clocks, optical frequency synthesis or low noise microwave
generation by frequency division. In terms of soliton dynamics, this
demonstration constitutes the first measurement of the carrier envelope
offset frequency of a temporal dissipative Kerr cavity soliton. This opens a new route to studying  nonlinear dynamics of solitons. A further important step in the future will be to phase lock both $f_{rep}$ and $f_{0}$ to an external radio frequency reference, and thereby achieve a phase stabilized self-referenced microresonator frequency comb. This can be achieved by controlling both $f_{rep}$ and $f_{0}$ independently
via changing the pump frequency detuning along with either the pump
power \cite{DelHaye2008} or by applying a stress to the resonator
with a piezoelectric crystal \cite{Papp2012}. The crystalline microresonator
based microwave to optical link can be made more compact with almost
all optical components being fiber optic based, and aside from the
optical microresonator the non-fiber based components (filters and
sum frequency generation stages) can be replaced with fiber based
components in the future. Finally, the external broadening stage itself,
required in the present case, can in suitably dispersion engineered
optical microresonators be avoided, when making use of soliton induced
higher order spectral broadening \cite{Coen2013,Brasch2014,Okawachi2011}. 
Eventually, this provides a path to counting the cycles of light using chipscale microresonators.

\section*{Funding Information}
This work was supported
by a Marie Curie IIF (J. D. J), the Swiss National Science Foundation (T.  H.), the European Space
Agency (V. B), a Marie Curie IEF (C.
L.), the Eurostars program, and the Defense Advanced Research Program
Agency (DARPA) PULSE program, grant number W31P4Q-13-1-0016. 

\section*{Acknowledgments}
{$^{\dagger}$}These authors contributed equally to this work.

\bibliography{Jost_s3}


\end{document}